# Discharge-produced plasma extreme ultraviolet (EUV) source and ultra high vacuum chamber for studying EUV-induced processes

A Dolgov[1], O Yakushev[2,3], A Abrikosov[2,4], E Snegirev[2,3], V M Krivtsun[2,3], C J Lee[1] and F Bijkerk[1]

[1] MESA+ Institute for Nanotechnology, University of Twente, Enschede, The Netherlands
[2] RnD–ISAN ltd. / EUVLabs ltd., Troitsk, Moscow, Russia
[3] Institute for Spectroscopy, Moscow, Russia
[4] Moscow Institute of Physics and Technology, Dolgoprudny, Moscow reg., Russia

E-mail: a.dolgov@utwente.nl

**Abstract.** An experimental setup that directly reproduces Extreme UV-lithography relevant conditions for detailed component exposure tests is described. The EUV setup includes a pulsed plasma radiation source, operating at 13.5 nm; a debris mitigation system; collection and filtering optics; and an UHV experimental chamber, equipped with optical and plasma diagnostics. The first results, identifying the physical parameters and evolution of EUV-induced plasmas are presented. Finally, the applicability and accuracy of the *in situ* diagnostics is briefly discussed.

# 1. Introduction

The next generation of photolithography, extreme ultraviolet (EUV) lithography, is expected to be required at the 13.5 nm node [1,2]. The ionizing photon flux, and vacuum requirements create a challenging operating environment. One of the most important requirements is optics purity with most of the optical elements expected to last for the lifetime of the photolithography tool. The basic optical element in EUV lithography (EUVL) is the multilayer mirror (MLM). MLMs consist of approximately 50 bi-layers of Mo/Si that are 6.7 nm thick, with the uppermost layer being a protective layer that is 1.5-2 nm thick. A thorough review of these mirrors is given in [3]. Multilayer mirrors in EUVL are expensive, high-technology items, making it desirable to extend their useful lifetime as much as possible.

Previous research has demonstrated that MLMs may lose their reflectivity due to their surfaces becoming contaminated with amorphous carbon or surface oxidation, induced by intense EUV radiation [4,5]. One solution that is currently under investigation is using EUV-induced plasma for in-line cleaning [6]. Experiments have shown that carbon etching can be achieved under certain conditions, but the physics and chemistry of the etching process is still not fully understood. One of the reasons for this lack of understanding is that the characteristics of the EUV radiation-induced plasma are poorly known.

In this article we present an experimental setup that allows EUV-induced processes to be studied. These processes include carbon contamination, surface oxidation, and plasma cleaning, as well as plasma-assisted processes like ion sputtering and blistering [7, 8, 9]. Carbon contamination was previously studied using an electron beam as a proxy for EUV [10]. However, electron beam irradiation is not fully analogous to the situation found in an EUV tool, where high energy photons play a significant role in plasma chemistry. This problem can be solved by using synchrotron radiation [11,12], but such experiments are impossible in ordinary laboratories. Currently there are a number of laboratory EUV sources, based on vacuum discharge in xenon or tin. The former sources are more common, but the latter have much greater conversion efficiency (up to 2% in tin compared to 0.5-0.7% in xenon) [1].

For the study of elementary processes in the EUV-induced plasma, the apparatus must be equipped with the appropriate diagnostics. Recent direct studies of the EUV-induced plasma were successfully carried out by using a method called microwave cavity resonance spectroscopy [13], but in this case, surface processes, such as secondary electron emission and surface plasma formation cannot be studied. However, EUV induced secondary electrons, which escape from the surface of the topmost or capping layer, are able to trigger surface reactions, possibly with an even higher efficiency than direct EUV photoionization. These secondary electrons produce additional ionization in the gas above the optic's surface, leading to a plasma sheath with an increased degree of ionization. Such a near-surface layer with enhanced electron density may strongly influence the surface photo- and plasma-chemistry, including practical cleaning schemes.

As we can see, a lot of physical phenomena (photoionization, secondary electron emission, ion plasma production etc.) contribute to the processes above the surface under EUV radiation. At the moment it is not clear which of them are more efficient at cleaning surfaces, despite the fact that the influence of EUV light on cleaning has been shown experimentally in ref [6]. However, near-surface elementary processes, occurring in the EUV-induced plasma, are poorly understood. Depending on the type of a gas and pressure in the volume above the surface, different types of ions and radicals (cleaning agents) predominate at different times. The formation of different cleaning agents strongly depends on the speed of plasma transport and the decay time of the plasma. Therefore, an experimental setup for the study of time evolution of the plasma parameters is required to understand EUV-induced surface plasma processing.

An experimental setup, that is able to characterize low-density EUV-induced plasmas using a time-resolved Langmuir probe system is presented in this work. Below, the radiation source, with intensity and pulse duration relevant to EUVL, and diagnostic tools for experimental measurements, are described in detail.

The EUV source, based on a discharge-produced plasma in tin vapor, irradiates samples in a clean vacuum chamber. The latter is designed for the investigation of surface and volume physical and chemical processes, induced or assisted by EUV. The benefits of this setup are high EUV power density, technical simplicity, and good control over the background gas and electric fields in the clean cylindrically symmetrical chamber. The addition of surface biasing and cylindrically symmetric fields makes the setup amenable to numerical calculations.

We also describe the diagnostic equipment for studying the EUV-induced plasma. Such plasma is difficult to study because of the low density (the electron density hardly exceeds $10^7$-$10^{10}$ cm$^{-3}$). In such cases optical diagnostics are inefficient, while previous work suggested that Langmuir probes would be infeasible [14]. Here, however, we show that Langmuir probes for measuring the plasma density and temperature are effective. We succeeded in obtaining the temporal dynamics of the electron temperature and density in a low-density EUV-induced plasma. These measurements are presented at the end of the article.

2.Setup
The installation shown in figure 1 (PROTO 2 setup) consists of several main parts. The first part is the discharge vacuum chamber, equipped with spouts for the entry of laser radiation and the installation of diagnostic sensors. In addition, the chamber has four flanges available for optical observations of the discharge. The chamber pumping system consists of two pumps, capable of 1000 l/s, resulting in an operating pressure of $10^{-3}$ Pa.

The lower part of this chamber contains the source of EUV radiation—highly ionized tin plasma—initiated by a laser pulse in the discharge gap (3-4 mm) between the anode and cathode. The anode and liquid tin-coated cathode rotate synchronously at a frequency of up to 30 Hz and operate at a potential difference up to 5 kV. One load of tin is sufficient to maintain continuous operation for 17 hours, which is approximately 100 MShots.

The tin is evaporated from the cathode using an Nd:YAG laser that generates pulses with a duration of 50 ns, and energies up to 37 mJ, at a pulse repetition rate of 1.6 kHz. The average laser radiation power incident on the cathode is approximately 56 watts.

The upper part of the discharge chamber contains collecting optics and the debris mitigation system. The collection optics and debris mitigation system are mounted in an independent housing, that is attached to the side-wall of the upper discharge chamber at an angle of 30° relative to the plane of the electrodes. The radiation is introduced into the chamber using a metal conical light guide with 10 mm gap

To prevent neutral tin atoms from entering the guide, hydrogen flows in the opposite direction with a flow rate of 0.001 l/s. Along the guide, there are six SmCo magnets, that create a field of 0.5 T in the guiding structure. The magnetic field extends over a length of 10 cm, and protects the collector mirrors from tin ions with energies up to 100 KeV. Finally, a rotating foil trap is used to protect the collection optics from droplets of tin created by the discharge [15]. The foil trap consists of 150 molybdenum foil blades, 20 mm long, on a disk with a diameter of 150 mm. The trap rotates with the frequency of 100 Hz and is located directly in front of the light collector. To pass through the foil trap, particles are required to have a speed of 500 m/s. Combined with the magnetic field trap, only Z-pinch ions above 150 keV are incident on the collector optics.

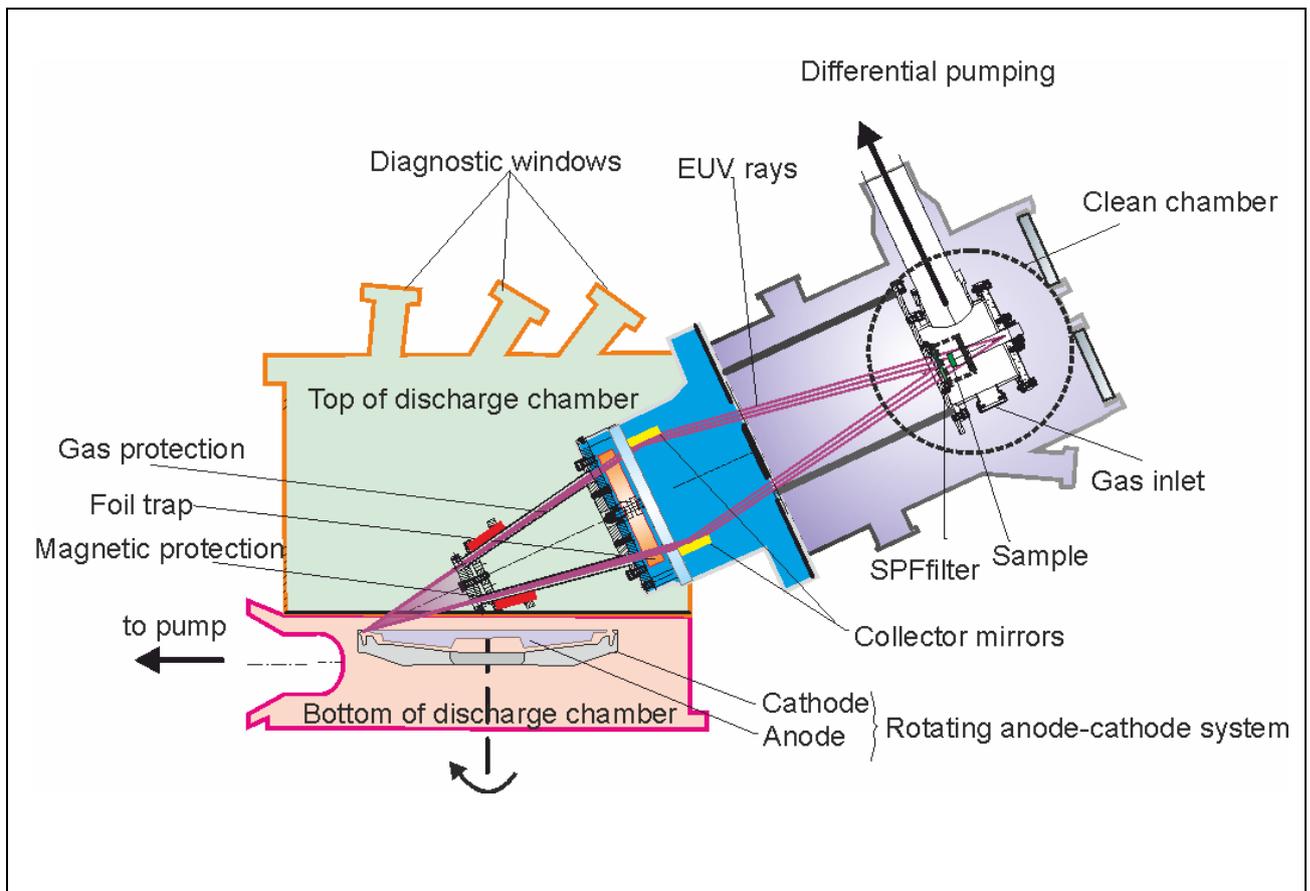

Figure. 1. EUV source, based on discharge-produced plasma in tin vapor, with a UHV vacuum chamber for the investigation of surface and volume physical and chemical processes, induced or assisted by EUV.

The light collector consists of grazing incidence cylindrical mirrors that are coaxially mounted. The mirror substrates are made from sital, and shaped by deep grinding and polishing. The radius of the mirrors is 75 mm, and the height is 45 mm. The focal distance of the collector is 480 mm. The mirror coating consists of a 100 nm thick layer of Mo, deposited by [some technique here] and has an EUV reflectivity of 83% with surface roughness ~ 7 nm.

To prevent carbon from accumulating on the mirror surface [4], the mirrors are biased at 150 V [6]. This bias voltage provides a flux of low energy hydrogen ions to the surface, which act as a cleaning agent.

*2.1.UHV chamber*
EUV radiation from the tin plasma is refocused by the collector optics at the sample location in the clean chamber. The UHV chamber is separated from the main volume by a 200 nm thick Zr filter, mounted on a nickel grid [16]. The spectral purity filter (SPF) is opaque from the deep UV to visible range and transmits EUV radiation at a wavelength of the 13.5 nm. Radiation spectra from Z-pinch discharge with and without SPF in range 10-25 nm is presented in figure 2.

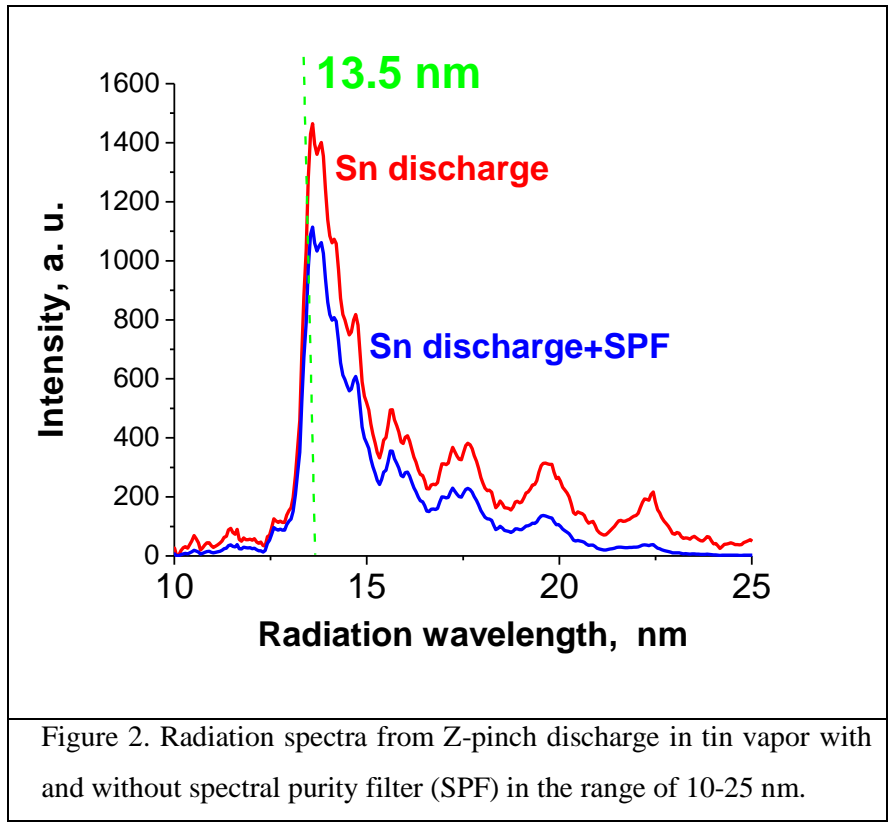

Figure 2. Radiation spectra from Z-pinch discharge in tin vapor with and without spectral purity filter (SPF) in the range of 10-25 nm.

The filter is mounted to the chamber using indium seals, and provides sufficient physical strength to allow isolation even in the presence of large differential pressures. This allows the pressure in the clean chamber to be varied from $10^{-6}$ Pa to 100 Pa. The maximum difference in pressure was limited by two factors: the tensile strength of the spectral filter, and the quality of the filter's seal. At maximum pressure, the flow from the clean chamber to the discharge chamber was sufficient to influence the source operation.

The UHV chamber (see figure 3) is equipped with heating, cooling—including a cryogenic dosing system—differential pumping, and sample mounts with the provision for biasing samples.

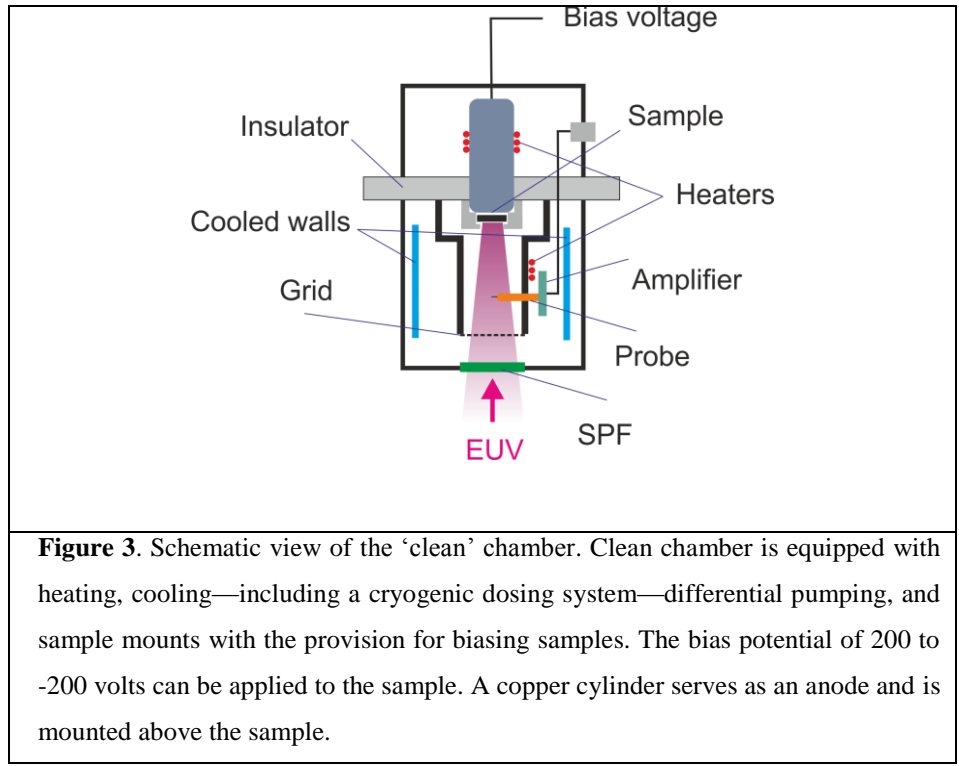

**Figure 3**. Schematic view of the 'clean' chamber. Clean chamber is equipped with heating, cooling—including a cryogenic dosing system—differential pumping, and sample mounts with the provision for biasing samples. The bias potential of 200 to -200 volts can be applied to the sample. A copper cylinder serves as an anode and is mounted above the sample.

The energy characteristics of EUV radiation of discharge plasma were measured using the pin diode (AXUV 100, IRD Inc.), shielded by an SPF filter. The power of EUV radiation, introduced to the 'clean' chamber, was measured using a specially designed and calibrated calorimeter, based on an AD590 temperature sensor.

The basic parameters of the EUV source and sample chamber are provided in the table 1.

**Table 1.** Basic parameters of the PROTO2 tin EUV source with exposure chamber.

| Main parameters of PROTO 2 experimental setup | |
|---|---|
| Maximum source power consumption | 3 kW |
| Mean electric energy of discharge output | 1.85 J/pulse. |
| In-band discharge to optical conversion ($2\pi$ solid angle) | 1.5% |
| Pulse repetition rate | 1.6 kHz |
| Pulse duration of EUV radiation | 100 ns |
| EUV intensity in focus spot D=6 mm without SPF | 0,75 W/cm$^2$ (4.9 10$^{-4}$ J/cm$^2$/pulse) |
| EUV intensity in focus spot D=6 mm with SPF | 0,13 W/cm$^2$ (0.85 10$^{-4}$ J/cm$^2$/pulse) |
| **Exposure clean chamber and probe system parameters** | |
| Chamber temperature range | from -100$^\circ$C to + 150$^\circ$C |
| Exposure chamber pressure range | from 10$^{-6}$ Pa to 100 Pa |
| Sample bias voltage range | from -200V to 200 V |
| Langmuir probe bias voltage range | from -190V to 190 V |

### 3. Study of the EUV-induced plasma.

To characterise the low-density EUV-induced plasma that forms in the clean chamber, a system of Langmuir probes and low noise electronics was installed.

*3.1. Probe measurement scheme.*

Figure 4 provides a full scheme of probe characteristics measurement of the EUV-induced plasma. There are three mounting holes for Langmuir probes (figure shows only one probe) along the axis of the cylinder of the anode. The distances from the probe locations to the sample surface are 12, 20, and 28 mm. Since each probe, in principle, disturbs the plasma, all measurements are performed with a single probe installed at a chosen location to minimize the probe influence while still obtaining an accurate measurement of the spatial dependence of the plasma. The probe consists of a steel wire of 0.5 mm diameter in a ceramic shell that leaves a 5 mm length of wire exposed to the plasma.

Before entering the chamber, the chamber, gases are passes through a nitric trap to get rid of impurities. The temperature of the sample and the anode is measured using termopare and are kept at a constant temperature through active cooling and heating. Heating is provided by cement resistor heaters. Cooling is achieved using liquid nitrogen. The electrical connections pass through a sealed iron housing that provides screening from external crosstalk. In case of experiments, that described in this paper the clean chamber was filled with hydrogen.

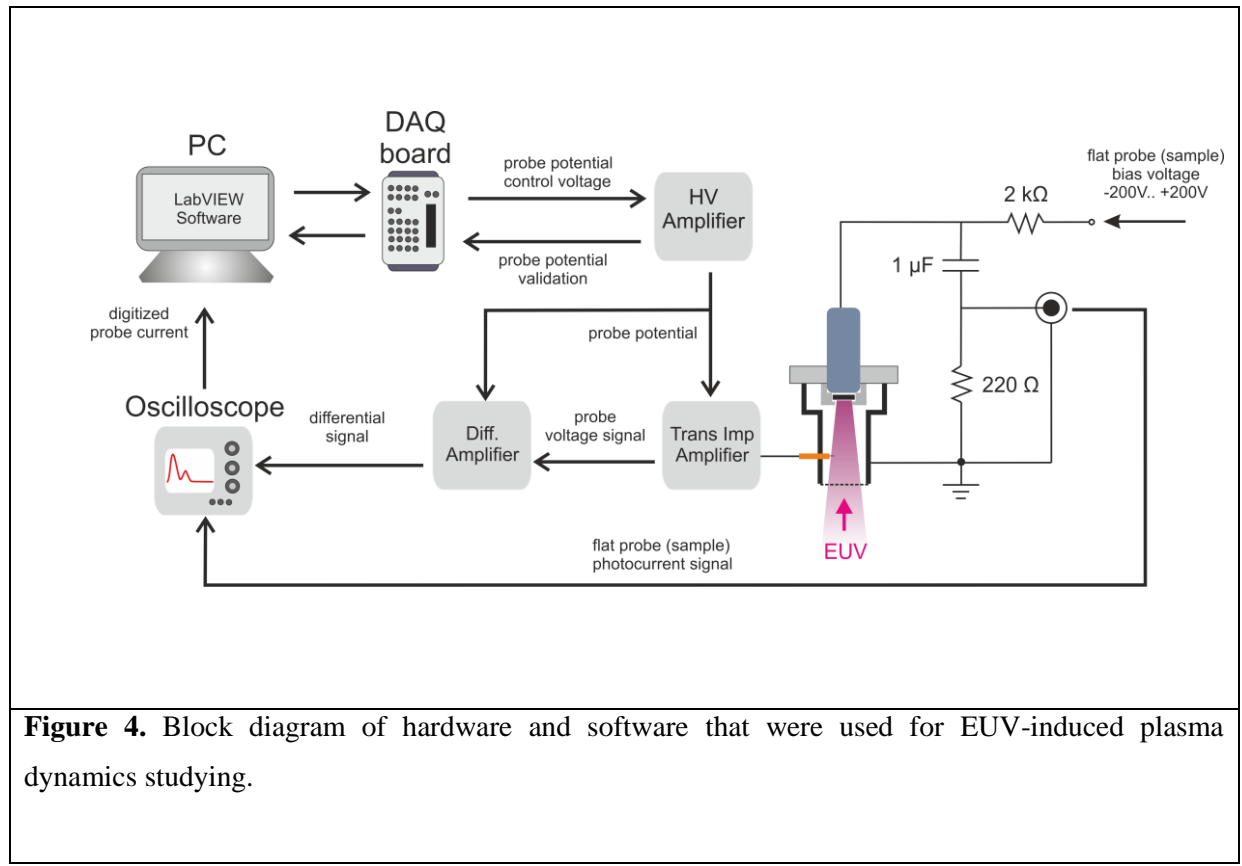

**Figure 4.** Block diagram of hardware and software that were used for EUV-induced plasma dynamics studying.

A low-noise amplifier circuit was used to acquire the probe signals. It consists of two amplifier cascades. The current signal of the probe is converted to a voltage signal by a transimpedance amplifier, based on an ОУ OPA656. The non-inverted input of the amplifier, as well as the power supply terminals, have the same potential as the output of the high-voltage amplifier that generates the bias voltage. Due to the feedback of the transimpedance amplifier, the inverting input, to which the probe is connected, has the same voltage. To minimize crosstalk, this amplifier is mounted in the clean chamber next to the probe. It must be mentioned that, under typical experimental high vacuum conditions, no fabric-reinforced laminate may be used because it releases gases. Thus, all the boards placed in the clean chamber are ceramic-based. The signal from the transimpedance amplifier is transferred to a differential amplifier that removes the common-mode interference from the signal and amplifies the signal by a factor of five.

The resulting signal is recorded by an oscilloscope (Tektronix DPO 3014), which, to reduce noise, operates in averaging mode. When averaging 512 oscilloscope records, a noise amplitude of approximately 10 nA is reached, which is low enough to measure the currents due to a low-density plasma.

Figure 4 also provides a scheme for measuring the current through the sample, which behaves as if it is a flat probe. The focus of the collector contains a sample holder (a large copper electrode with an attachment for inch-size silicon substrates). A bias potential of 200 to -200 volts can be applied to the sample, using an external power source. A copper cylinder, which serves as an anode, is mounted above the holder. The holder and the anode are included in the circuit provided in the figure 4. This circuit allows real-time monitoring of the current leaving the sample, when exposed to EUV radiation.

**4. Experimental results**

The photocurrent measurements of the sample (copper disk) were carried out at different hydrogen gas pressures in the clean chamber. The pressure was fixed for each of the individual measurements. When a chosen bias value was reached, the time dependence of the current in the circuit between the sample and the cylindrical anode was recorded. In a vacuum this current is characterized by a duration comparable to the duration of EUV pulse radiation, i.e., ~ 100 ns.

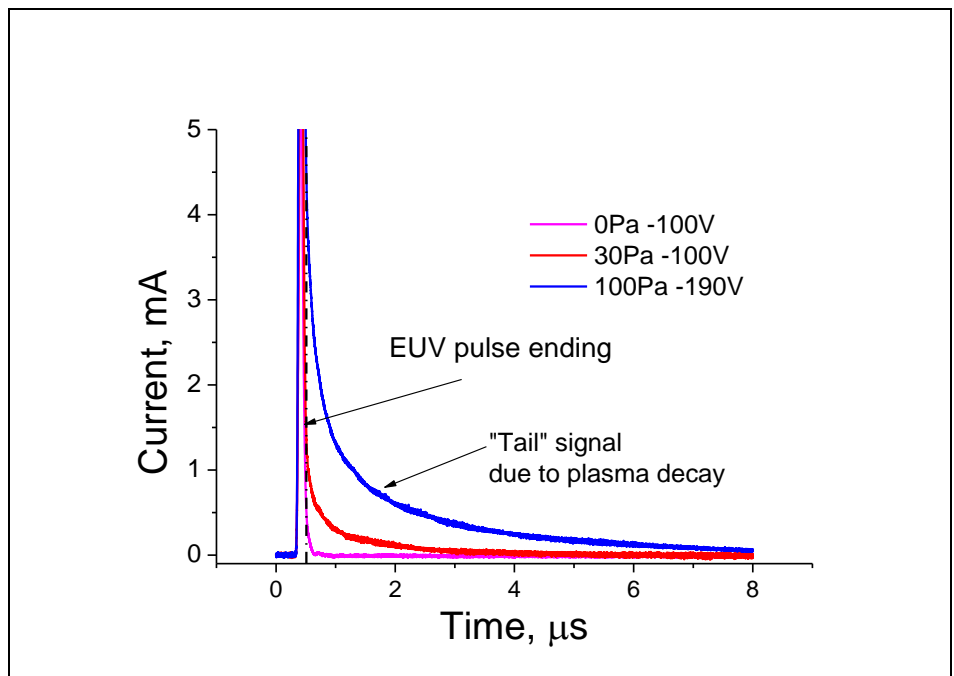

**Figure 5**. The photocurrent, measured at different gas pressures and bias voltages. The tail is due to the ion current and illustrates the lifetime of the plasma formed during photo-ionization, and the ionization of the gas by photoelectrons close to the sample.

In the presence of a gas, the signal has a long plasma tail (see figure 5), fading in the 4-6 μs after the start of acquisition (and long after the end of the EUV pulse). The slow tail is due to the ion current— illustrating the lifetime of the plasma formed during photoionization—as well as due to the ionization of the gas by photoelectrons above the sample. The integration of the slowly decaying tail provides a charge that can be used to determine the ion dose incident on the sample. Moreover, assuming that plasma is localized in the volume of the cylindrical anode, it is possible to provide an approximate estimate of ion density. For this purpose, the pulse is integrated from the end of the EUV radiation pulse (moment $t_0$):

$$n_e = n_i \approx \frac{2}{eV_a} \int_{t_0}^{t_{max}} I_P(t)dt$$

,

where $V_a$ is the volume of the cylinder of the anode, $I_P$ is the current of the sample at gas pressure *P, e* is the electron charge, $t_0$ is the time of the end of the EUV pulse. The multiplier 2 takes into account the fact that the charge is collected on the sample from the near-electrode layer and not from the entire volume.

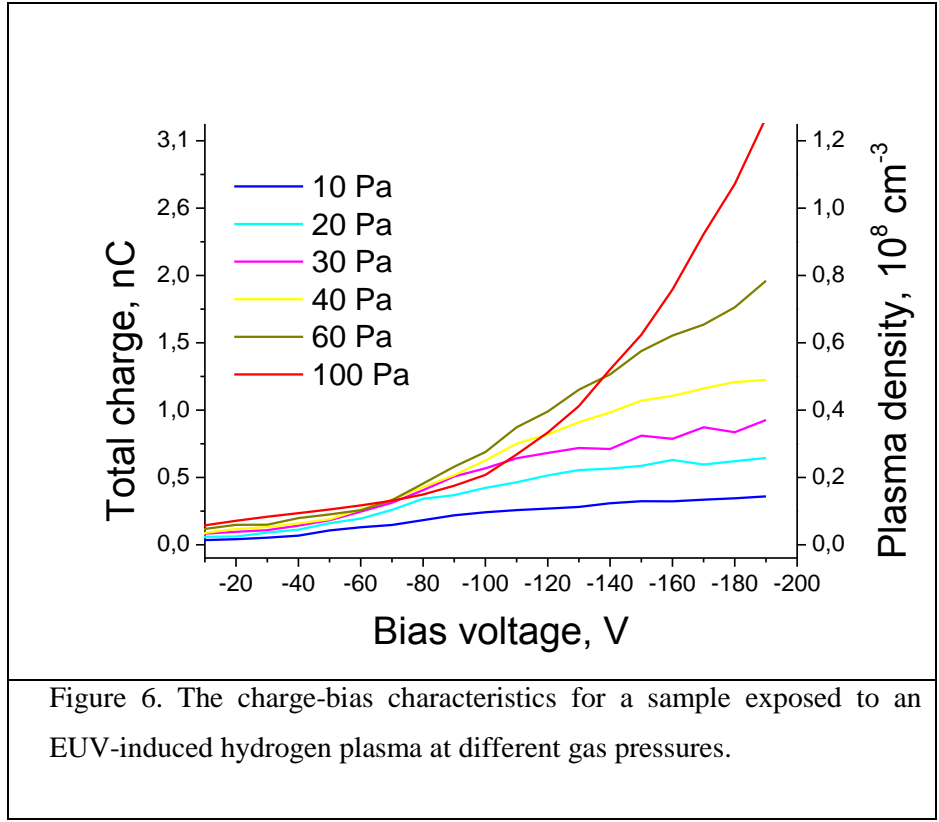

Figure 6. The charge-bias characteristics for a sample exposed to an EUV-induced hydrogen plasma at different gas pressures.

Figure 6 shows the charge-bias characteristics for a sample exposed to an EUV-induced hydrogen plasma. In this case, based on the curve in figure 6, the plasma ion density is estimated to be $n_i \sim 10^7$ cm$^{-3}$. Thus, as expected, the density of the plasma is low.

### *4.1. Probe measurements*
At the time of each measurement, the pressure, and the offset voltage supplied to the sample, over which the probe is placed, remain unchanged. Measurements were taken at relatively high pressure (20 Pa), and high sample potential (20 V) to limit the effect of the Langmuir probe on the plasma. The potential of the probe is controllably varied, and the current pulse from the probe was acquired for each probe potential. Cutting the current data at a constant time results in a *I-V* characteristics. Figure 8 shows the *I-V* curve for a number of different times.

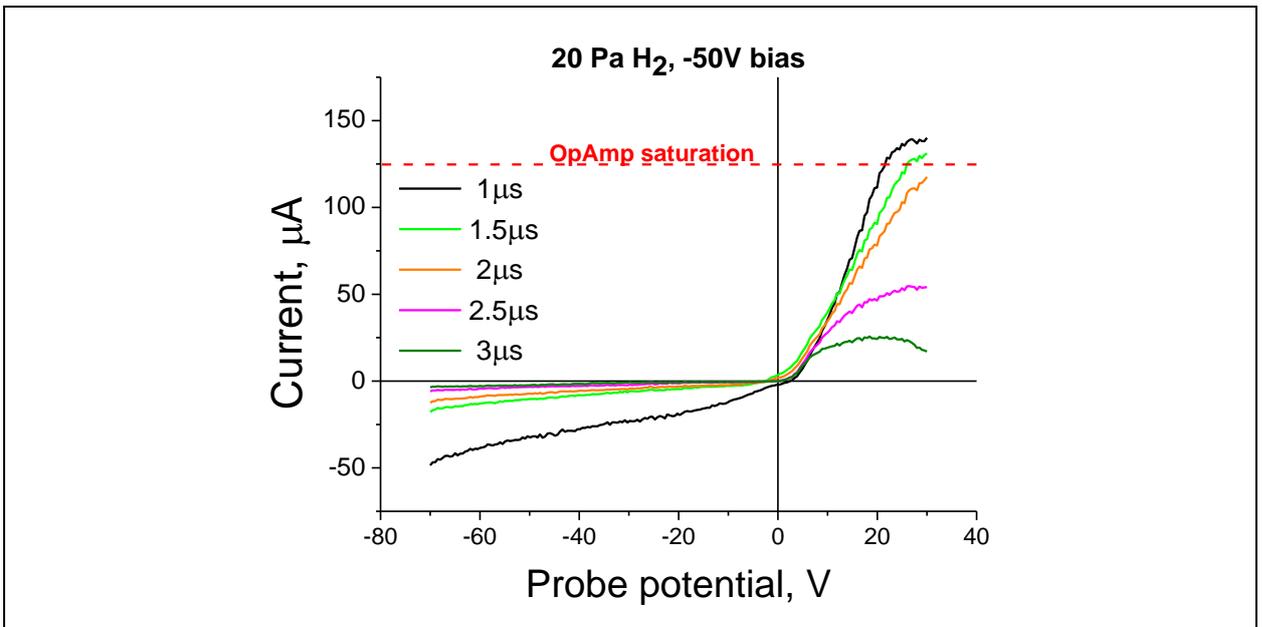

Figure 7. Volt-Ampere characteristics of the EUV induced plasma measured at the different times for a gas pressure of 20 Pa, and a sample bias of 50 V. $T_e$ can be estimated from the slope of the near-exponentially decreasing portion of the curve. The electron contribution to the current appears abruptly, which indicates that the plasma is, as expected, relatively cold.

The Volt-Ampere characteristics of the EUV induced plasma, shown in figure 7, are quite typical for a low-density, low-temperature plasma. The floating potential of the probe, $V_f$, varied within a range of ±2 V. The electron contribution to the current appears abruptly, which indicates that the plasma is, as expected, relatively cold ($T_e \sim$ 1-2 eV). To estimate the density of electrons, Laframboise's theory of charge collection [17,18] was applied (see figure 8). ). The estimated electron density was $3\text{-}4\cdot 10^7$ cm$^{-3}$, which is in good agreement with the estimate based on the charge-bias characteristic made above. Taking into account electro neutrality of the plasma, we know densitydensity of ions at the same time, which is the density of active cleaning species after EUV pulse. As we can see at figure 8 immediately after the radiation pulse, plasma temperature is relatively high due to hot photoelectrons. After first two microseconds electrons cool to a temperature of about 1-1.5 eV, which is typical for low-temperature plasmas. This temperature is maintained in the plasma for up to 4 microseconds, or during the entire plasma lifetime. To study plasma dynamics during the EUV pulse, it is necessary to take into account the emission from the surface of the probe, which we have not yet studied.

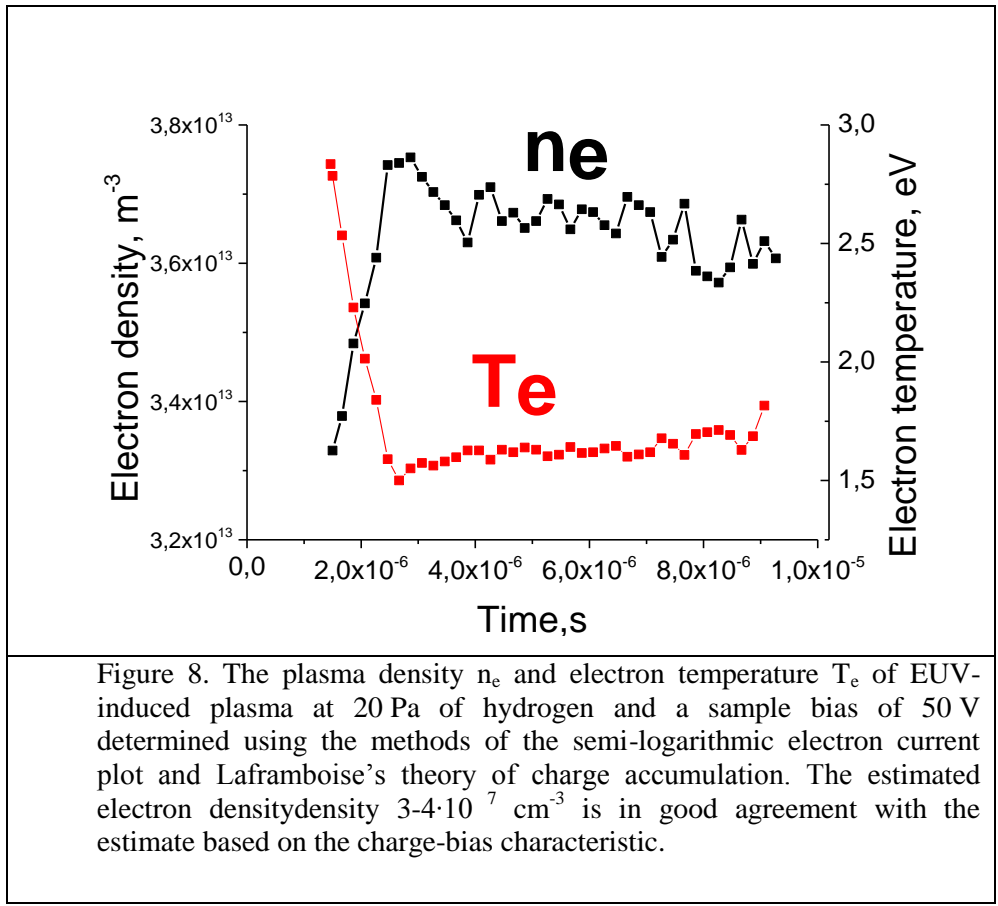

Figure 8. The plasma density $n_e$ and electron temperature $T_e$ of EUV-induced plasma at 20 Pa of hydrogen and a sample bias of 50 V determined using the methods of the semi-logarithmic electron current plot and Laframboise's theory of charge accumulation. The estimated electron densitydensity $3\text{-}4\cdot 10^7$ cm$^{-3}$ is in good agreement with the estimate based on the charge-bias characteristic.

## 5. Conclusion

An experimental setup for the study of EUV-induced plasma directly in time of plasma-surface interactions at Extreme UV-lithography relevant conditions has been described. The installation is equipped with a system of Langmuir probes that are able of measuring the temporal and spatial development of very low density plasmas. Since the surface chemistry of MLMs is often driven by plasma-induced process, knowledge of the plasma properties is critical to understanding surface chemistry. The experimental setup described here is expected to allow fundamental studies of surface physics and chemistry to be informed by accurate plasma characterization.

The determined plasma characteristics include: the temperature of electrons, and the density of plasma. Measurements can be performed over a wide range of pressures and cathode biases. It is also possible to study the temporal dynamics of plasmas after the EUV pulse.

Our initial results show that the characteristic development of a low-temperature plasma, excited by pulsed EUV radiation, are in good agreement with preliminary estimates of the plasma parameters. The experiments show that the electron density for EUV-induced plasma increases from the moment of the beginning of the pulse and lasts a few microseconds. In these measurements the density of all gas pressures studied never exceeds $5\cdot 10^8$ cm$^{-3}$ and drops after 3 microseconds.

**Acknowledgments**
This work is part of the research program "Controlling photon and plasma induced processes at EUV optical surfaces (CP3E)" of the "Stichting voor Fundamenteel Onderzoek der Materie (FOM)" which is financially supported by the Nederlandse Organisatie voor Wetenschappelijk Onderzoek (NWO). The CP3E programme is cofinanced by Carl Zeiss SMT GmbH (Oberkochen), ASML (Veldhoven), and the AgentschapNL through the Catrene EXEPT program.

**References**
[1]     Bakshi V 2006 *EUV sources for lithography*. (SPIE press).
[2]     Silverman P J. 2005 *Journal of Micro/Nanolithography, MEMS, and MOEMS* **4.1** 011006-011006.
[3]     Louis E, Yakshin A E, Tsarfati T, Bijkerk F 2011 *Prog. Surf. Sci*
[4]     Chen J Q, Louis E, Lee C J, Wormeester H, Kunze R, Schmidt H, Schneider D, Moors R, van Schaik W, Lubomska M, and Bijkerk F 2009 *Optics Express* **17** 16969-16979.
[5]     Hiroaki O et al. 2007*Japanese Journal of Applied Physics* **46** L633–L635
[6]     Dolgov A., et al. 2014 *J. Phys. D: Appl. Phys* **47** 065205.
[7]     Allain J P, Hassanein A, Allain M M C, Heuser B J, Nieto M et al 2006 *Nuclear Instruments and Methods in Physics Research Section B: Beam Interactions with Materials and Atoms,* **242** 520-522
[8]     Kuznetsov A S, Gleeson M A, and Bijkerk F, *J. Phys. Condens. Matter* **24** 052203 (2012)
[9]     Kuznetsov A S, Gleeson M A,; van de Kruijs R W E and Bijkerk F 2011 *Proc. SPIE 8077, Damage to VUV, EUV, and X-ray Optics III*, 807713
[10]    Kurt R et al. 2002 *Proc. SPIE,* **4688** 0277
[11]    Socol Y, Kulipanov G N, Matveenko A N, Shevchenko O A, and Vinokurov N A 2011 *Phys. Rev. ST Accel. Beams* **14** 040702
[12]    Al-Montaser Bellah Al-Ajlony, Kanjilal A, Sivanandan S H and Hassanein A 2012 *J. Vac. Sci. Technol. B* **30** 041603
[13]    van der Horst R M et al 2014 *J. Phys. D: Appl. Phys*. **47** 302001
[14]    van der Velden H L *Radiation Generated Plasmas*. Diss. Ph. D, 2008.
[15]    Shmaenok L A, de Bruijn C C, Fledderus H F, Stuik R, Schmidt A A et al 1998
        *Proc. SPIE 3331, Emerging Lithographic Technologies II,* **90**
[16]    Chkhalo N I at el.2012 *J. Micro/Nanolith. MEMS MOEMS* **11** 021115
[17]    Chen F F.2003 *IEEE-ICOPS Meeting, Jeju, Korea*.
[18]    Laframboise J G 1966 *Univ. Toronto Aerospace Studies Report* **11**